\documentclass[twocolumn,showpacs,graphicx,amsmath,amssymb]{revtex4}
\usepackage{graphicx}

\begin{document}

\title{The production rate of the coarse grained
Gibbs entropy and the Kolmogorov-Sinai entropy: a real connection ?}  
\author{Massimo Falcioni} 
\affiliation{Dipartimento di Fisica and Istituto dei Sistemi Complessi
del Consiglio Nazionale delle Ricerche, Universit\`a di Roma "La
Sapienza", P.le A.Moro 2, Rome 00185, Italy}
\author{Luigi Palatella} 
\affiliation{Dipartimento di Fisica and Istituto dei Sistemi Complessi
del Consiglio Nazionale delle Ricerche, Universit\`a di Roma "La
Sapienza", P.le A.Moro 2, Rome 00185, Italy}
\author{Angelo Vulpiani}
\affiliation{Dipartimento di Fisica and Istituto dei Sistemi Complessi
del Consiglio Nazionale delle Ricerche, Universit\`a di Roma "La
Sapienza", P.le A.Moro 2, Rome 00185, Italy}
\affiliation{INFN, Sezione di Roma ``La Sapienza''}

\begin{abstract}
We discuss the connection between the Kolmogorov-Sinai entropy,
$h_{KS}$, and the production rate of the coarse grained Gibbs entropy,
$r_G$. Detailed numerical computations show that the (often accepted)
identification of the two quantities does not hold in systems with
intermittent behavior and/or very different characteristic times and
in systems presenting pseudo-chaos. The basic reason of this fact is
in the asymptotic (with respect to time) nature of $h_{KS}$, while
$r_G$ is a quantity related to short time features of a system.

\end{abstract}


\maketitle

\section{Introduction}
The physical entropy is a quantity that plays a key role in the
understanding of the basic laws ruling the macroscopic behavior of
systems with many degrees of freedom. We just mention the Boltzmann's
microscopic interpretation of the macroscopic Clausius equilibrium
entropy, and the celebrated H-theorem \cite{A}.

On the other hand the term entropy is widely used also in contexts
different from thermodynamics and statistical mechanics. In the
information theory there is the Shannon entropy
\cite{shannon,kolmogorov,gwh(e)}, while in dynamical systems one uses
the Kolmogorov-Sinai entropy (and other entropic quantities like, {\em
e.g.}, the Renyi entropies) \cite{gwh(e),boffetta}.

The connection between some properties of the non-equilibrium
thermodynamical systems and the underlying chaotic dynamics, recently,
has attracted the interest of many scientists \cite{B,C,D,E}. The main
question is if and how the dynamical characteristic quantities (such
as Lyapunov exponents or Kolmogorov-Sinai entropy) are related to
macroscopic physical properties ({\em e.g.}: diffusion coefficients
and entropy production rate) \cite{F,G,B,C,D,E,PB1,PB2,AP,rosa}.

Some authors, on the basis of reasonable arguments and numerical
computations on simple dynamical systems, claim the existence of a
relation between the Kolmogorov-Sinai entropy, $h_{KS}$, and the
production rate, $r_G$, of a suitably averaged coarse-grained Gibbs
entropy \cite{latora}.

The main goal of this work is to show that this is not the generic
case. To this end we will study two different kinds of discrete
time systems:

\begin{itemize}

\item[a)] 
one-dimensional intermittent maps, with local Lyapunov exponents very
different from the mean value;

\item[b)]
slightly coupled maps with very different (uncoupled) Lyapunov
exponents.
\end{itemize}

We will show that when there are different (local Lyapunov exponents)
time scales in the different regions of the phase space, there may be
no room for an identification, on a meaningful time interval, between
$h_{KS}$ and $r_G$. This is so because, when averaging {\em at a fixed
time} the entropy contributions originating from regions with
different time-scales, densities with different space-scales get mixed,
at possibly different relaxation stages.

A second point we want to stress is that, even if a linearly
increasing time behavior of the coarse-grained entropy is observed,
the rate of growth is not necessarily given by the Kolmogorov-Sinai
invariant of the system. This will be shown by studying discretized
versions of a chaotic map (i.e. an automaton). In such a case, the
memory of the chaotic character of the original system may allow for a
{\em transient} chaotic-like regime long enough to make the behavior
of the strictly periodic system indistinguishable from the one of its
truly chaotic ancestor.  That is, pseudochaos is at work \cite{H,I}:
the long time properties of the system, like $h_{KS}$($=0$), remain
hidden and do not affect quantities, like $r_G$($\neq 0$), related to
short time features of the system.

The paper is organized as follows. In Sect.~2 we recall some basic
concepts and methods in chaotic dynamical systems and statistical
mechanics. In addition we give a simple argument for the connection
between $h_{KS}$ and $r_G$, stressing the weak points of the 
argument. In Sect.~3 we discuss the results of numerical computation
for the time evolution of the coarse-grained Gibbs entropy in 
systems with ``non trivial'' dynamical features, {\em i.e.} with
intermittency and different characteristic times. Sect.~4 is 
devoted to conclusions and discussions.

\section{A brief overview of basic facts}

Since the pioneering work of Kolmogorov \cite{kolmogorov,boffetta},
the relevance of the Kolmogorov-Sinai entropy for a proper
characterization of the behavior of a chaotic dynamical system was
clear. To perform the computation of $h_{KS}$ one has to choose a
partition ${\cal A}$ of the phase space and to assign each cell of the
partition an integer value $i$.  In such a way, the trajectories of a
dynamical system, with continuous states, sampled at discrete times,
\[
{\bf x}(1), {\bf x}(2),...,{\bf x}(j),..., {\bf x}(T)
\]
become symbolic sequences 
\begin{equation}\label{simbolica}
i(1),i(2),...,i(j),...,i(T)
\end{equation}
whose meaning is that at time $j$ the trajectory is in the cell
labeled by $i(j)$. Then one defines the probability of each word (or
block) of length $n$, $p^{({\cal A})}(k_1,k_2,..,k_n)$, counting how
many times one meets the word $k_1,k_2,...,k_n$ along the sequence
(\ref{simbolica}). The entropy of the blocks of size $n$, $H^{({\cal
A})}(n)$, thus reads
\begin{equation}\label{block}
H^{({\cal A})}(n) = - \sum\limits_{k_1,..,k_n} 
p^{({\cal A})}(k_1,..,k_n) 
\log p^{({\cal A})}(k_1,..,k_n).
\end{equation}
The Kolmogorov-Sinai entropy is defined by the $\sup$ over all
partitions of the asymptotic value of the rate of increase of
$H^{({\cal A})}(n)$, i.e.
\begin{equation}\label{kolmentro}
h_{KS} = \sup\limits_{{\cal A}} \lim\limits_{n \rightarrow \infty} 
\frac{H^{({\cal A})}(n)}{n} \, .
\end{equation}
The quantity $h_{KS}$ is a numerical invariant that gives a good
characterization of a chaotic system, but unfortunately it is almost
impossible to compute analytically (except very few simple cases) and
also rather difficult from a numerical point a view. From a physical
point of view it is rather natural to use regular partitions with
hyper-cubic cells of edge $\epsilon$. Let us denote with
$H^{(\epsilon)}(n)$ the $n$-block entropy on a partition of this kind
and $\epsilon$-entropy the limit
\begin{equation}\label{epsentro}
h(\epsilon) = \lim\limits_{n \rightarrow \infty} 
\frac{H^{(\epsilon)}(n)}{n} = \lim\limits_{n \rightarrow \infty}
H^{(\epsilon)}(n+1) - H^{(\epsilon)}(n) \, \, .
\end{equation}
It is a remarkable fact that $h(\epsilon)$, computed with different
values of $\epsilon$, can give very interesting information about the
properties of the system\cite{gwh(e),boffetta}. Moreover it is
possible to obtain $h_{KS}$ by considering the limit $\epsilon \to 0$
in (\ref{epsentro}), instead of the $\sup$ operation in
(\ref{kolmentro}). By recalling the Shannon-McMillan equipartition
theorem (stating that, for large $n$, the number of ``typical''
$n$-words increases as $\exp[h(\epsilon)\, n]$) one has that
$h(\epsilon)$ gives the (asymptotic in time) exponentially growing
rate of the number of typical trajectories of the system, in the limit
of high resolution (as measured by $\epsilon$). It is intuitive that
this growing rate must be linked to the exponentially fast separation
of nearby trajectories. The Pesin theorem is the rigorous statement of
this idea: all the expanding directions contribute to the
diversification of the trajectories and to the increase of their
number
\begin{equation}\label{pesin}
h_{KS} = \sum\limits_{\lambda_i > 0} \lambda_i \,\, ,
\end{equation}
where ($\lambda_i>0$) means the sum over positive $\lambda_i$. Pesin's
identity (\ref{pesin}) provides us with a useful alternative way to
compute $h_{KS}$. The Lyapunov exponents can be numerically computed
without particular difficulties, even in high-dimensional systems; on
the contrary, because of the exponential proliferation of the
$n$-words, the Kolmogorov-Sinai entropy becomes rapidly unattainable
by numerical methods (a part low-dimensional systems).  Therefore very
often the Pesin formula is basically the unique way to compute
$h_{KS}$.

Notice that $h_{KS}$ is an entropy rate defined on the ensemble of the
trajectories of a system (according to some stationary probability
measure), even if, by means of ergodicity, it is practically computed
using only one single long trajectory. Qualitative arguments may be
given \cite{zaslavsky} to support a connection between $h_{KS}$ and
the rate of variation of an entropy-like quantity, that we call Gibbs
entropy, defined on the phase space of the system. A possible line of
reasoning is the following.

Consider a deterministic dynamical law
\begin{equation}\label{contin} 
{\bf x} \to T^t  {\bf x} 
\end{equation}
(where ${\bf x}$ is a $D$-dimensional vector) and a probability
density $\rho({\bf x},t)$, that gives a distribution of states of
the system throughout its phase space at a time $t$.  We define the
Gibbs entropy of $\rho$ as follows
\begin{equation}\label{def-S} 
S(\rho_t)= - \int \rho({\bf x},t) \ln [\rho({\bf x},t)] d{\bf x} \,\, , 
\end{equation}
{\em i.e.} its conditional entropy with respect to a uniform density.
For chaotic dissipative systems, where $\rho({\bf x},t)$ tends to a
singular (fractal) measure, definition (\ref{def-S}) becomes
meaningless. Nevertheless, following a nice idea of Ruelle \cite{L},
one can avoid this difficulty, simply by adding (or considering
unavoidably present) a small noise term in the evolution law: in such
a way one obtains a $\rho({\bf x},t)$ continuous with respect to the
Lebesgue measure.  If $J({\bf x},t)$ is the Jacobian of
(\ref{contin}), a straightforward computation gives
\begin{equation}\label{grow} 
S(\rho_t)=S(\rho_0)+ 
\int \rho({\bf x},t) \ln |J({\bf x},t)| d{\bf x} \,\, . 
\end{equation}
In the case of volume conserving evolutions, one has:
$S(\rho_t)=S(\rho_0)$. To allow for an entropy variation one needs a
coarse-graining. Let us consider a hyper-cubic partition, as introduced
above, and let us define the probability $P^{\epsilon}(i, t)$ to find
the state of the system in the cell $i$ at time $t$:
\begin{equation}\label{prob}
P^{\epsilon}(i, t)= \int_{\Lambda_i^{\epsilon}}
 \rho({\bf x},t) d{\bf x} 
\end{equation}
where $\Lambda_i^{\epsilon}$ is the region singled out by the $i-$th
cell. Let us introduce the $\epsilon$-coarse-grained Gibbs entropy
\[
S^{\epsilon}(P_t) = - \sum_i P^{\epsilon}(i, t) \ln P^{\epsilon}(i, t).
\]
If $\epsilon$ is small enough $S(\rho_t)$ and $S^{\epsilon}(P_t)$
are trivially related:
\[
S^{\epsilon}(P_t) \simeq S(\rho_t) + D \ln \left( \frac{1}{\epsilon} \right).
\]
If one considers a distribution of initial conditions that is
different from zero only over one (or very few) cell(s), 
one has, for ${\epsilon}$ small enough and a time not too short,
\begin{equation}\label{cg-entro}
S^{\epsilon}(P_t)=S^{\epsilon}(P_0) + h_{KS} t \, .
\end{equation}
One can argue as follows. Assume that the system has $m$ positive
Lyapunov exponents and that $\rho({\bf x},0)$ is localized around
${\bf x}^c(0)$. In a suitable reference system (with the axes along
the eigendirections of the Lyapunov exponents), if $\rho({\bf x},0)$
has a Gaussian shape, $\rho({\bf x},t)$, for some times, is still well
approximated by a Gaussian with variances
\begin{equation}\label{sigma}
\sigma_j^2(t)=\sigma_j^2(0) \exp\{ 2 \lambda_j t \}
\end{equation}
therefore:
\begin{equation}\label{rho}
\rho({\bf x},t) \simeq \prod_{j=1}^D 
{1 \over \sqrt{2 \pi \sigma_j^2(t) }}
 e^{ - {{(x_j -x_j^c(t))^2} \over {2 \sigma_j^2(t)}}}
\end{equation}
where ${\bf x}^c(t)$ is the state evolved from ${\bf x}^c(0)$.
From this, in the non-grained case, one gets
\[ 
S(\rho_t)=S(\rho_0)+  \sum_j \ln { \sigma_j(t) \over \sigma_j(0)}
=S(\rho_0) +  \sum_{j=1}^D  \lambda_j t \,\, .  
\]
It is clear that $S(\rho_t)=S(\rho_0)$ if the phase space volume is
conserved.  Considering now the coarse-graining (\ref{prob}), one has
that along the directions of the negative Lyapunov exponents ($m+1,
m+2, \dots $), for a long enough $t$:
\[ \sigma_k(t) \sim \sigma_k(0) e^{- |\lambda_k | t} \le \epsilon 
\]
This implies that
\[ P_i^{\epsilon}(t) \simeq 
\prod_{j=1}^m 
{1 \over \sqrt{2 \pi \sigma_j^2(t) }}
 e^{ - {{(x_j^{(i)} -x_j^c(t))^2} \over {2 \sigma_j^2(t)}}}
\] 
and therefore
\[ S^{\epsilon}(P_t)=S^{\epsilon}(P_0) +  
\sum_{j=1}^m \lambda_j t \,\, . 
\] 
With the aid of the Pesin's formula (\ref{pesin}) eq.~(\ref{cg-entro})
follows. Let us stress that the transition from (\ref{grow}) to
(\ref{cg-entro}) is allowed by the fact that, in the presence of a
coarse-graining, the contracting eigendirections (corresponding to the
negative values of the Lyapunov exponents) cannot balance the effects
of the expanding ones.

At this point we have to notice that, by definition, the Gibbs entropy
(\ref{def-S}) explicitly depends on the particular chosen initial
density. In the discussion here above this dependence may be labeled
by the cell, ${\bf x}^c$, where the distribution is initially
different from zero. On the contrary, $h_{KS}$ is an asymptotic global
property of the system. So, one may expect that a density independent
behavior, as in (\ref{cg-entro}), can be found only in ``friendly''
dynamical systems, {\em i.e.} systems with no fluctuations. A generic
system possesses a certain degree of intermittency, so that, for
instance, the expanding and contracting properties may strongly depend
on the phase space region the trajectory is visiting. This calls for
an averaging over the initial condition ${\bf x}^c$, weighted, say, by
the natural invariant measure of the system:
\begin{equation}\label{average}
S(\rho_t) \to S(t) =\int S(t \vert {\bf x}^c) 
\rho_{eq}({\bf x}^c) d {\bf x}^c
\end{equation}
where $S(t|\mathbf{x_c})$ is $S(\rho_t)$ with $\rho_0(\mathbf{x})$
localized around $\mathbf{x_c}$.  The same averaging procedure leads to
$S^{\epsilon}(t)$ from $S^{\epsilon}(P_t)$. This operation yields
intrinsic quantities, that can depend on global properties of the
system.  The interesting question is whether the simple relation
(\ref{cg-entro}) survives, as an observable property, for the averaged
coarse-grained entropy. Notice that equation (\ref{cg-entro}), besides
its conceptual interest, would result in a numerically simple way to
determine $h_{KS}$.  In some systems suitable ranges of $t$ and
$\epsilon$ exist where the relation is verified
\cite{latora}. However, we believe, in agreement with Ref.~\cite{rosa},
that what has been found in Ref.~\cite{latora} is just a ``lucky''
coincidence. With regard to that, it is important to stress that in
the above presented arguments, for the derivation of (\ref{cg-entro}),
there are (at least) two delicate points:
\begin{itemize}
\item[a)]
both Lyapunov exponents and Kolmogorov-Sinai entropy are 
quantities defined in the limits of high resolution 
($\epsilon \to 0$) and long times ($t \to \infty$);
\item[b)]
a behavior like (\ref{sigma}) holds only for ``short'' time,
{\em i.e.}: 
\[
t \alt \frac{1}{\lambda_{1}} \ln \frac{1}{\sigma(0)} \,\, ;
\]
and this is so also for the linear behavior in (\ref{cg-entro})
that is expected to be valid for 
\[
t \alt \frac{1}{\lambda_{1}} \ln \frac{1}{\epsilon}.
\]
If intermittency is present one has to replace $\lambda_{1}$ with
the largest local Lyapunov exponent. 
\end{itemize}
Since for a coarse-grained Gibbs entropy a finite $\epsilon$ is
mandatory, it is not obvious that the previous time regimes a) and b)
have a non-empty overlap. One may note that in the entropic analysis
of the n-words one of the two asymptotic limits can be relaxed:
i.e. one can work with non-infinitesimal $\epsilon$ and therefore
obtain the $\epsilon$-entropy $h(\epsilon)$ which is an asymptotic in
time quantity associated with a finite tolerance $\epsilon$. However,
once the size of the cells for the coarse-grained Gibbs entropy is
fixed, nonetheless $\rho(x,t)$ evolves developing structures on scales
$l(t) \approx \epsilon \exp (\lambda t)$ increasing in time. Therefore
it is not trivial at all that a simple relation between
$dS^{\epsilon}(t)/dt$ and $h(\epsilon)$ exists.

\section{A numerical study of simple dynamical systems.}

In this section we present an entropic analysis of simple dynamical
systems which show, in spite of their low dimensionality, non trivial
features.

From a practical point of view the computation of $S^{\epsilon}(t)$ 
is performed as follows:

\begin{enumerate}

\item
we select several starting conditions 
${\bf x}^{j}_1(0)$, ${\bf x}^{j}_2(0)$, $...$, ${\bf x}^{j}_N(0)$ 
all located in the $j$-th box of size $\epsilon^D$

\item
we let evolve all the $N \gg 1$ starting conditions up to a time $t$ obtaining
${\bf x}^{j}_1(t),{\bf x}^{j}_2(t),...,{\bf x}^{j}_N(t)$

\item
we calculate 
\[ p^{\epsilon,j}(k,t) = 
\frac{1}{N} \sum\limits_{i=1}^N \delta({\bf x}^j_i(t),k)
\] 
with 
\[
\delta({\bf x},k) = \left \{
\begin{array}{ll}
1 & \textrm{ if } {\bf x} \in \Lambda_k^{\epsilon} \\
0 & \textrm{ otherwise }
\end{array}
\right.
\]

\item
we compute the entropy of  $p^{\epsilon,j}(k,t)$ defined as
\begin{equation}
S^{\epsilon}(j,t) = 
- \sum\limits_{k=1}^N p^{\epsilon,j}(k,t) \log p^{\epsilon,j}(k,t)
\end{equation}

\item 
we average this quantity on the coarse-grained invariant distribution
$p_{eq}(j)$ obtained from $\rho_{eq}({\bf x})$ with the procedure of
Eq.(\ref{prob}).  Thus we have
\begin{equation}\label{gibbs}
S^{\epsilon}(t) = 
\sum\limits_{j=1}^N p_{eq}(j) S^{\epsilon}(j,t). 
\end{equation}


\end{enumerate}

\subsection{Intermittent map}

A recent paper \cite{rosa} shows that the value of $r_G$ in the
Manneville map \cite{manneville} significantly differs from the value
of $h_{KS}$. These authors perform the analytical calculation of
$S(t)$ starting from only one particular condition, namely the box
containing the point $x=0$, that in the Manneville map is the point
with the lowest local Lyapunov exponent. Since in this map the value
of the local Lyapunov exponent ranges from very low values
($\lambda(x) \rightarrow 0$ for $x\rightarrow 0$) to values
considerably grater than $1$, the authors correctly argue that the
discrepancy is due to the variability of $\lambda(x)$. Nevertheless
the authors of \cite{rosa} do not perform the final average over the
initial condition. In the following we see that the averaging
procedure (\ref{gibbs}) is not able to recover the condition of
Eq. (\ref{cg-entro}) neither in the Manneville map nor in a
less-intermittent map. Let us start with a simple case, i.e. the
modified tent map given by
\begin{equation}\label{tenda}
x_{t+1} = \left \{
\begin{array}{lll}
x_t/p & \textrm{if} & x_t<p \\
(1-x_t)/(1-p) & \textrm{if} & p<x_t<1. 
\end{array} \right.
\end{equation}
This map is the standard tent map if $p=1/2$, while for 
small values of $p$ one has an intermittent behavior
characterized by two very different local Lyapunov exponents,
namely:
\begin{equation}
\lambda_+ = -\log p,\:\: \lambda_- = -\log (1-p). 
\end{equation} 
The stationary distribution is constant between $0$ and $1$
and consequently we have
\begin{equation}\label{h_K}
h_{KS} = p \lambda_+ + (1-p) \lambda_- = -p \log p -(1-p) \log (1-p)
\end{equation}
assuming its maximum value for $p=1/2$ and decreasing its value for $p
\rightarrow 0$. Note that this map presents an intermittent behavior,
but not in a critical way like in the Manneville's map, and the
distribution of the length of the ``laminar'' zones (the permanence in
the zone $[p,1]$) is simply exponential without power law tail.
Nevertheless we see that already in this system Eq.(\ref{cg-entro})
does not hold.

\begin{figure}[!b]
\includegraphics[width=8.3 cm]{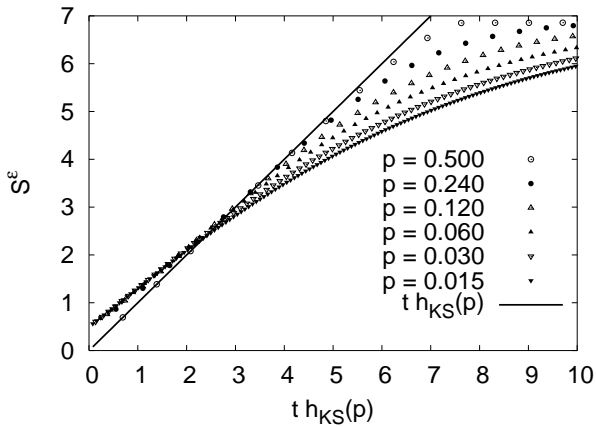}
\caption{\label{fig1} $S^{\epsilon}(t)$ as a function of $t h_{KS}(p)$ with 
$\epsilon = 10^{-3}$ and different values of $p$. Note that 
Eq.(\ref{cg-entro})
should give the straight line $t h_{KS}(p)$ for all values of $p$. }
\end{figure}

The numerical results are shown in fig.\ref{fig1}. We set $\epsilon =
10^{-3}$ and we scale the time axes with the inverse of $h_{KS}(p)$, given
by Eq.(\ref{h_K}), so if Eq.(\ref{cg-entro}) holds we should observe
that all the curves, with different $p$, collapse on the straight
line $S^{\epsilon}(t) = h_{KS}(p) t$. As one can see the agreement is
good only for $p=1/2$ (with no intermittency) while becomes worse and
worse decreasing $p$.  The main point is that the linear behavior of
$S^{\epsilon}(t)$ should hold, according to the heuristic arguments of
Ref.~\cite{latora}, till to a time given by
\begin{equation}
t_{lin} \simeq \frac{1}{\lambda_1} \log \left ( \frac{1}{\epsilon} \right ) 
\end{equation}
corresponding, on the scaled time of Fig.\ref{fig1}, to the value $-
\log \epsilon \simeq 7$.  This is observed for $p=1/2$ but for small
values of $p$ one has that $S^{\epsilon}(t)$ increases in time with a
``wrong'' slope (different from $h_{KS}$) and later it exhibits a
rather long crossover.

The origin of this effect is in the intermittent behavior of the
system.  Indeed the realizations $S^{\epsilon}(j,t)$ starting in the
zone $[0,p]$ are spread on the whole interval $[0,1]$ after few steps
almost reaching the asymptotic value of $ -\log \epsilon$ while the
``realizations'' starting for example near the unstable equilibrium
point $x = 1/(2-p)$ takes several time steps to reach the saturation
giving a dominant contribution to the rate of increase of
$S^{\epsilon}(t)$. In this way the $S^{\epsilon}(t)$ computed with
Eq.(\ref{gibbs}) does not increase in time following the naive
argument yielding to Eq.(\ref{cg-entro}). The reason of the
discontinuity in $S^{\epsilon}(1)$ for $p \rightarrow 1/2$ is
explained in Appendix A.

Fig.\ref{fig2} shows other interesting properties of the behavior of
$S^{\epsilon}(t)$ for $p=0.1$ and $p=0.4$ at varying the value of
$\epsilon$. As one can see, in the slightly intermittent case $p=0.4$
the rescaled curves collapse together confirming the assumption of
Eq.(\ref{cg-entro}); while in the intermittent case $p=0.1$ the curves
do not collapse and only for very low values of $\epsilon$ a linear
growth of $S^{\epsilon}(t)$ is present. Let us stress the fact that,
at variance with the case shown in Fig. \ref{fig2}b, in the
intermittent case (Fig.\ref{fig2}a) the crossover regime (after the
linear one and before the saturation) is very long and is comparable
with the duration of the linear regime.



\begin{figure}[!h]
a)\includegraphics[width=8.3 cm]{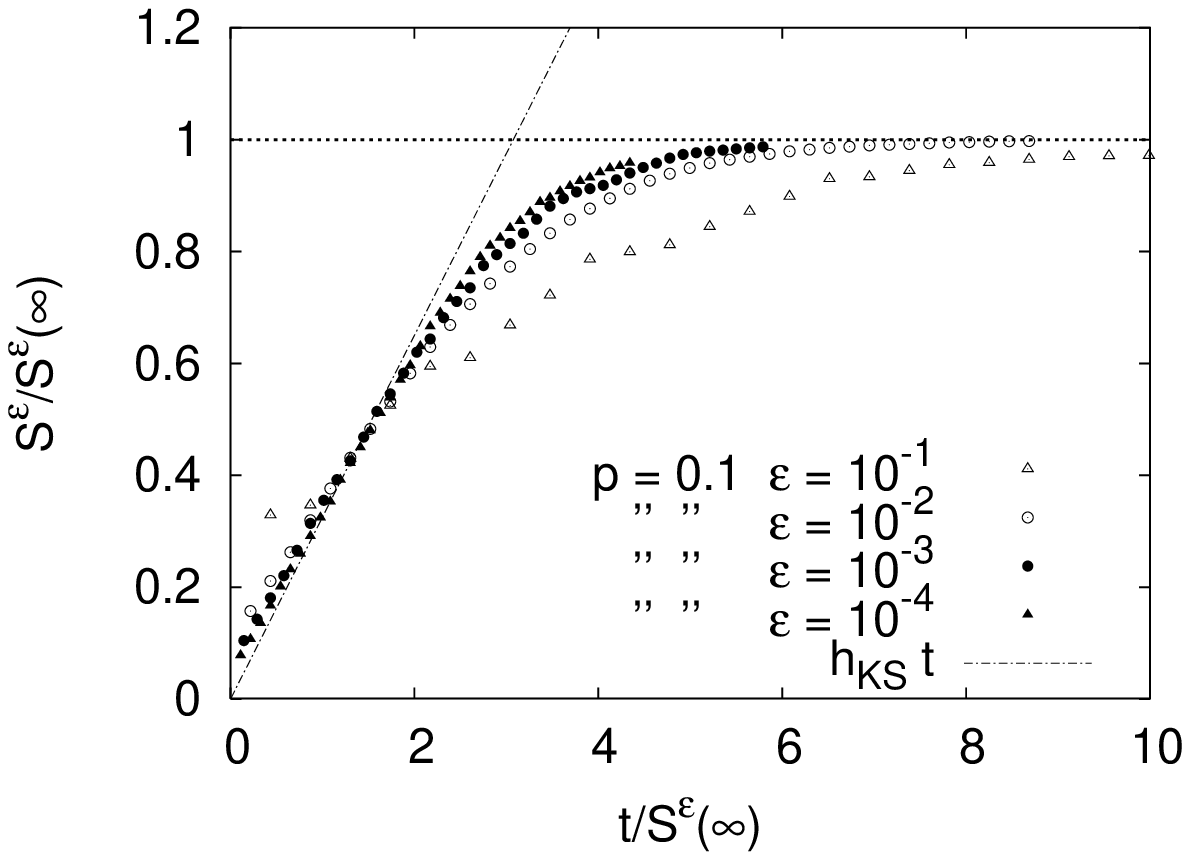}\\
b)\includegraphics[width=8.3 cm]{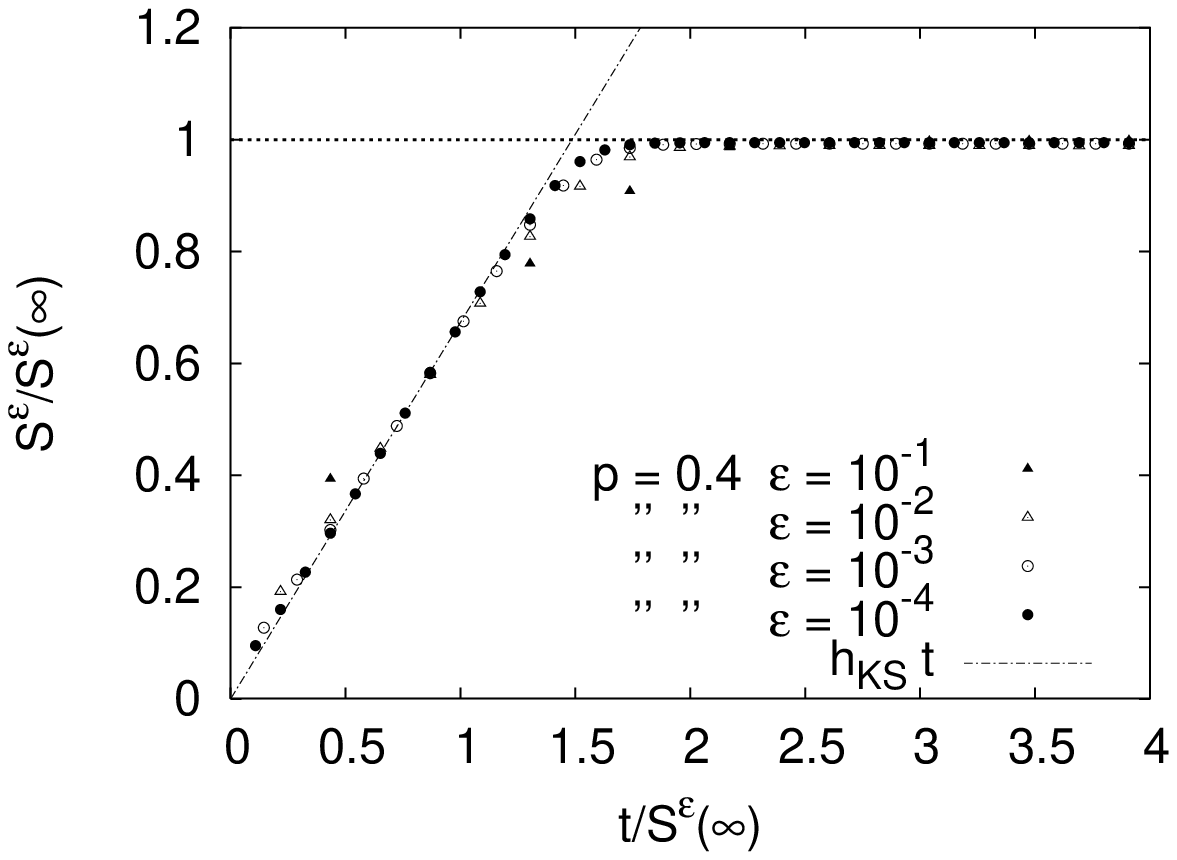}
\caption{\label{fig2} $S^{\epsilon}(t)/S^{\epsilon}(\infty)$ as a function 
of $t /S^{\epsilon}(\infty)$ with different values of
$\epsilon$. In fig. a) $p=0.1$, in fig. b) $p=0.4$ }
\end{figure}

To connect our results to the ones of Ref.~\cite{rosa} we also perform
the same calculation for the Manneville map given by
\begin{equation}\label{manneville}
x_{t+1} = \left \{
\begin{array}{lll}
x_t + k x_t^{z} & \textrm{if} & x_t<d \\
(1-x_t)/(1-d) & \textrm{if} & d<x_t<1. 
\end{array} \right.
\end{equation} 
where $d$ fulfills
\[
d + kd^{z} = 1.
\]
In the range $3/2<z<2$ and $k \ll 1$ there is an invariant
distribution \cite{gwh(e),rosa}. Under these conditions the
distribution of the permanence time $\tau$ has a power law tail with
$\langle \tau \rangle <\infty $ while $\langle \tau^2 \rangle$
diverges.  The only difference from the tent map calculation is that
the invariant distribution $\rho_{eq}(x)$ is not constant and
therefore we have to compute it numerically. Also in this case we
observe a ``wrong'' slope (lower than $h_{KS}$) and an extremely long
crossover behavior.

\subsection{2D map}

Let us now discuss a system with two different time-scales. 
We choose the following 2D maps for the variable $x_t$ and $y_t$
\begin{equation}
\left \{
\begin{array}{ll}
x_{t+1} = & f(x_t) + \sigma \cos(2\pi(x_t+y_t))\\
y_{t+1} = &  r y_t + \sigma \cos(2\pi (x_t+y_t)) \:\: {\rm mod}\: 1
\end{array} \right.
\end{equation}
with
\begin{equation}
f(x) = \left \{
\begin{array}{lll}
x/p & \textrm{if} & x<p \\
(1-x)/(1-p) & \textrm{if} & p<x<1  
\end{array} \right.
\end{equation}
and $r$ is an integer greater than $1$. The map for $x_t$ is a
modified tent map, like in the previous section, while the map for
$y_t$ is a generalized Bernoulli shift. We choose $p=0.3$ and $r=5$
in order to have two largely different uncoupled Lyapunov exponents,
namely $\lambda_x \simeq 0.61$ and $\lambda_y = \log 5 \simeq 1.61$.
We use this kind of coupling to avoid discontinuity at $x=0$ or $x=1$.

We will see that in this system the two Lyapunov exponents $\lambda_x$
and $\lambda_y$ have a role rather similar to that of $\lambda_+$ and
$\lambda_-$ for the system (\ref{tenda}).

\begin{figure}[!h]
\includegraphics[width=8.3 cm]{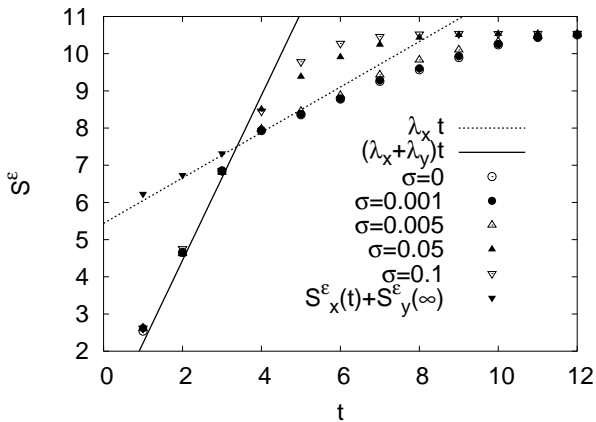}
\caption{\label{fig4} $S^{\epsilon}(t)$ as a function of $t$ with
different values of $\sigma$, $\epsilon = 5 \cdot 10^{-3}$, $p=0.3$, $r=5$}
\end{figure}

The coarse-grained Gibbs entropy of this 2D map is calculated starting
from a 2-dimensional cell of linear size $\epsilon$ and performing the
average over the two dimensional coarse-grained invariant
distribution, that also in this case turns out to be flat in the
slightly coupled case. We also numerically calculated the Lyapunov
exponents in the coupled case: for the small values of $\sigma$ used
in the numerical calculations, we observe no relevant changes from the
uncoupled case. We set $\epsilon =5 \cdot 10^{-3}$ and we study the
behavior of $S^{\epsilon}(t)$ with different values of $\sigma$, in
such a way we can study the crossover behavior at varying $\sigma$.
The results for are shown in fig.\ref{fig4}.

As one can see, at very small $t$, $S^{\epsilon}(t)$ increases with a
slope equal to the Kolmogorov-Sinai entropy $h_{KS} \simeq \lambda_x +
\lambda_y = -p \log p -(1-p) \log(1-p) +\log r$. After that the
phase space of the variable $y$ saturates and the slope changes. In
the case $\sigma<\epsilon$ the influence of the coupling is almost
negligible and the second slope is the same we should observe in the
1D case. In other words, as shown in the figure, we have
$S^{\epsilon}(t) \simeq S^{\epsilon}_y(\infty) + S^{\epsilon}_x(t)$
where $S^{\epsilon}_{x}(t)$ is the entropy calculated from the
marginal probability density $\rho_x(x,t) = \int {\rm d}y \rho(x,y,t)$.
In a similar way $S_y^{\epsilon}(t)$ is obtained from $\rho_y(y,t) =
\int {\rm d}x \rho(x,y,t)$.


When $\sigma \approx \epsilon$ the noise on the slowest
variable $x$ is enough to spread the distribution as fast as it
happens for the variable $y$ so no change in slope is observed.

Thus we observe that, for values of $\sigma >\epsilon$, $r_G$
corresponds quite well to the value of $h_{KS}$ in a broad time
interval while for values $\sigma \ll \epsilon$ this correspondence
longs only for a very short time. This is noteworthy because this map
has pratically the same $h_{KS}$ entropy at the different values of
$\sigma$ used (for $\sigma=0$ $h_{KS}=-p \log(p) - (1-p)\log(1-p) +
\log(r) = 2.22$ while for $\sigma=0.1$ we numerically obtain $h_{KS} =
2.19$) but with values of $\sigma > \epsilon$ we have $r_G \approx
h_{KS}$ while for $\sigma<\epsilon$ the equivalence is lost or longs
for a very short time.

It is interesting to compare Fig.\ref{fig4} with Fig.\ref{fig2}a: in
both cases the ``naive'' behavior (i.e. $S^{\epsilon}(t) -
S^{\epsilon}(0) = h_{KS}t$) may have a very short duration and there
is a long crossover. The origin of this crossover is due to a sort of
``contamination effect'' of different times (in other words different
mechanisms) involved. For the system (\ref{tenda}) this is due to
intermittency (the system ``feels'' $\lambda_+$ and $\lambda_-$) while
for the 2D map it is due to the existence of different times that are
relevant at different spatial resolution scales.

In systems with many degrees of freedom the ``contamination effect''
can produce rather impressive behaviors. As example we can cite the
case of fully developed turbulence where, because of the existence of
many different characteristic times, the growth of the distance
between two trajectories is a power law in time instead to be an
exponential (although the system is chaotic) \cite{M}.

\subsection{Discretized 1D map}

We now discuss a class of dynamical system with zero $h_{KS}$: the
discretized maps also called coarse-grained deterministic automata
\cite{PRL}.  We study the discretized (in phase space) version of 1D
map defined by the equation
\begin{equation}\label{tenda_discreta}
n_{t+1} = \eta \left \lfloor \frac{f(n_t)}{\eta} \right \rfloor 
\end{equation}
where $\eta$ is the discretization parameter (the number of
discretized state is $1/\eta$), $\lfloor .\rfloor$ denotes the integer
part and $f(x)$ is a tent map:
\begin{equation}
f(x) = \left \{
\begin{array}{lll}
x/p & \textrm{if} & x<p \\
(1-x)/(1-p) & \textrm{if} & p<x<1. 
\end{array} \right.
\end{equation}

\begin{figure}[!h]
\includegraphics[width=8.3 cm]{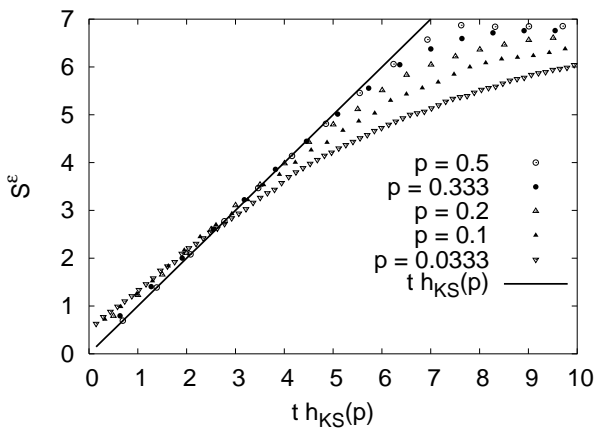}
\caption{\label{discreta} $S^{\epsilon}(t)$ as a function of $t h_{KS}(p)$ 
with different values of $p$, $\epsilon = 10^{-3}$, $1/\eta=3 \cdot 10^6$}
\end{figure}

Even if this system is the discretized version of a chaotic map the
finiteness of the available states forces the dynamics to become
periodic. In ref.\cite{PRL} it has been observed that the block
entropy $H^{\epsilon}(n)$ defined as in Eq.(\ref{block}) with
$\epsilon \gg \eta$ (i.e. several discrete states in each cell) for
small values of $n$ increases with a slope given by the Kolmogorov
entropy $h_{KS}$ of the corresponding continuous system
$x_{t+1}=f(x_t)$ while for
\begin{equation}
n \agt t_{p} \simeq - \frac{1}{h_{KS}} \log \eta
\end{equation} 
the block entropy stops increasing thus revealing the periodic nature
of the dynamics. We present the behavior of $S^{\epsilon}(t)$ in this
system, varying the value of $p$ in order to observe also the effect
of intermittency. As fig.\ref{discreta} shows, the coarse-grained
Gibbs'entropy increases as a function of the scaled time $t h_{KS}(p)$
with a slope comparable with $1$; while it reaches the saturation
values at scaled time given approximatively by
\[
t_{sat} \simeq  \frac{1}{h_{KS}} \log 
\left ( \frac{1}{\epsilon} \right ) \approx 7.
\]  
Note that, in spite of the fact that the
KS-entropy of (\ref{tenda_discreta}) is strictly zero, this behavior
is practically the same of the continous system, thus revealing one of
the main problem of the use of the coarse-grained Gibbs'entropy to
detect $h_{KS}$.  Indeed the block entropy $H^{\epsilon}(n)$``feels''
that the system is periodic only for $n>t_p$, but
\begin{equation}
h_{KS} t_{p} \simeq - \log \eta > -\log \epsilon. 
\end{equation}
The last inequality follows from the physical condition we use
i.e. that $\epsilon \gg \eta$ in order to have several state inside
each cell.  The main point is that since $\epsilon \gg \eta$
\[
t_p \simeq \frac{1}{h_{KS}} \log (1/\eta) 
\]
is larger than the saturation time $t_{sat}$. Therefore {\em the
coarse-grained Gibbs'entropy saturates well before the system feels to
be periodic}.

\section{Conclusions}

We have shown that, in spite of some folklore, there is a rather loose
relation between the Kolmogorov-Sinai entropy and the growth of the
coarse grained (Gibbs like) entropy.  Such claimed connection exists
only in very special cases, namely systems with a unique
characteristic time and very weak intermittency (i.e. small
fluctautions of the local Lyapunov exponent). On the contrary, in more
interesting (and closer to reality) systems, with multiple
characteristic times and/or non negligible intermittency, the relation
between $h_{KS}$ and $r_G$ holds (if any) only for a very short time.

The main reason of this is due to the asymptotic nature of $h_{KS}$
(as well as the $\epsilon$-entropy), i.e. its relevance at very large
time intervals. On the contrary the growth of the coarse grained
entropy only involves short time intervals, and during the early time
evolution of $S^{\epsilon}(t)$ one has entanglement of behaviours at
different characteristic space scales.

This phenomenon is rather similar to that observed in the spreading of
passive tracers in closed basins \cite{artale}. In such a case, if the
characteristic length scale of the Eulerian velocities is not very
small, compared with the size of the basin, both the diffusion
coefficient and the Lyapunov exponent do not give relevant information
about the mechanism of spreading.

We stress again that the failure of the relation between $h_{KS}$ and
$r_G$ is due to the fact that the growth of the coarse grained entropy
is not related to asymptotic properties (i.e. large time and small
resolution). This is particularly evident in deterministic discrete
states systems that, although non chaotic (i.e. with $h_{KS}=0$), show
a behaviour of $S^{\epsilon}(t)$ very similar to that observed in
genuine chaotic systems (with $h_{KS} \neq 0$).

\begin{acknowledgments}
We thank F. Cecconi, P. Grigolini and L. Rondoni for useful discussions.
\end{acknowledgments}

\appendix*

\section{A}

We discuss here the discontinuity in $S^{\epsilon}(1)$ for
$p \neq 1/2$ observed in fig.\ref{fig1}. Indeed for $p=1/2$ we have
$S^{\epsilon}(1)=\log 2$ while for $p = 1/2 - \xi$, with $\xi \ll 1$,
we have $S^{\epsilon}(1) \simeq 0.85$. Obviously $S^{\epsilon}(0)=0$
for all values of $p$. To understand the reason of this discontinuity
let us compute the value of $S^{\epsilon}(1)$.  If we start from the
j-th cell we have
\[
p^{\epsilon,j}(k,0) = \delta_{k,j}.
\]
If $p=1/2$ at time $1$ this cell will be spread on exactly two cells
and the trajectory starting at the border of a cell will go exactly on
the border of another cell so if, for sake of simplicity,
$j<1/2\epsilon$ (i.e. if the cell is in the region $x<1/2$), we obtain
\[
p^{\epsilon,j}(k,1) = \frac{1}{2}(\delta_{k,2j}+\delta_{k,2j+1})
\]
leading to $S^{\epsilon}(1)=\log 2$ for each $j$.
If we have $p=1/2-\xi$ with $\xi \ll 1$ the size of the phase space
region populated after one time step is essentially the
same. Nevertheless this time the trajectory starting at the border of
a cell does not always go to another cell border so in the averaging
procedure of Eq.(\ref{gibbs}) there will be some starting conditions
giving approximately a superposition of the cell borders after one
time step (for example $i=0$) and consequently a value of
$S^{\epsilon}(1)\approx\log 2$, but also some starting conditions $j$ 
resulting in
\[
p^{\epsilon,j}(k,1) = \frac{1}{2}
\delta_{k,2j+1}+\frac{1}{4} (\delta_{k,2j}+\delta_{k,2j+2}).
\]
These terms give a contribution to the average greater than $\log 2$ and
lead to the discontinuity of fig.\ref{fig1} when going from $p=1/2$ to
$p\neq 1/2$. Anyway this behavior is not important for greater values
of time so it does not influence the behavior of $S^{\epsilon}(t)$ for
$t > 1$.

In order to check our numerical results we also compute
$S^{\epsilon}(t)$ from $\rho(x,t)$ obtained by the Perron-Frobenius
equation.  In the case of the tent map we use directly the expression
of the formal evolution of $\rho(x,t)$ given by
\begin{equation}
\rho(x,t+1) =  p \rho(px,t) + (1-p)\rho(1-x(1-p),t)
\end{equation}
implemented in a recursive algorithm on a computer program.
We fix 
\[
\rho(x,0) = \left \{
\begin{array}{ll}
1/\epsilon & \textrm{ if } {\bf x} \in \Lambda_k^{\epsilon} \\
0 & \textrm{ otherwise }
\end{array}
\right.
\]
we let $\rho(x,t)$ evolves and then we compute $S^{\epsilon}(t)$.  The
results obtained with this method are in perfect agreement with those
in Section III and confirm the presence of the discontinuity.

\end{document}